\documentclass[aps,pra,twocolumn, showpacs]{revtex4-1}

\usepackage{amsmath,amssymb}
\usepackage{graphicx}
\usepackage{epsfig}
\usepackage{multirow}
\usepackage{color}
\newcommand{\element}[3]{\langle #1|#2|#3\rangle}

\newcommand{\ket}[1]{|#1\rangle}
\newcommand{\ThreeJ}[6]{\left(\begin{array}{ccc}
#1&#2&#3\\
#4&#5&#6\\
\end{array}\right)}
\newcommand{\SixJ}[6]{\left\{\begin{array}{ccc}
#1&#2&#3\\
#4&#5&#6\\
\end{array}\right\}}

\newcommand{\z}{\phantom{0}}
\newcommand{\vect}[1]{\vec{\boldsymbol{#1}}}

\begin{document}

\input{epsf}
% \include{epsf}
%\draft

\title{Rubidium Rydberg linear macrotrimers}
\author{Nolan Samboy$^{1,2}$}
\author{Robin C\^{o}t\'{e}$^1$}

\affiliation{$^1$Physics Department, University of Connecticut,
             2152 Hillside Rd., Storrs, CT 06269-3046}
\affiliation{$^2$Physics Department, College of the Holy Cross,
             1 College St., Worcester, MA 01610}

\date{\today}
\begin{abstract}

We investigate the interaction between three rubidium atoms in highly excited 
($58p$) Rydberg states lying along a common axis and calculate the potential energy surfaces 
(PES) between the three atoms. We find that three-body long-range potential wells exist in some of 
these surfaces, indicating the existence of very extended bound states that we label
\textit{macrotrimers}. We calculate the lowest vibrational eigenmodes and the resulting energy levels and 
show that the corresponding vibrational periods are rapid enough to be detected spectroscopically.
\end{abstract}

\pacs{03.65.Sq, 31.50.Df, 32.80.Ee, 34.20.Cf 
%32.80.Rm, 03.67.Lx, 32.80.Pj, 34.20.Cf
}

\maketitle
%%%%%%%%%%%%%%%%%%%%%%%%%%%%%%%%%%%%%%
\section{Introduction}
\label{sec:intro}
Ultracold Rydberg systems are a particularly interesting avenue of study. Translationally,
the atoms are very slow, yet their internal energies are very high. The large
excitation of a single electron leads to exaggerated atomic properties, such as long lifetimes, 
large cross sections, and very large polarizabilities~\cite{Gallagher}, which can lead
to strong interactions between Rydberg atoms~\cite{Anderson,Mourachko}. Such interactions
have led to various applications in quantum information 
processes over the past decade (see~\cite{Saffman-RMP} for a comprehensive review).

Another active area of research with Rydberg atoms is in the area of 
long-range ``exotic molecules''. Such examples include the \textit{trilobite} 
and \textit{butterfly} states, so-called because of the resemblence of their 
respective wave functions to these creatures. First predicted in~\cite{trilobites}, 
these quantum states were detected more recently in~\cite{pfau}. 
Also of interest are the formation and detection of macroscopic Rydberg molecules. 
In~\cite{macro-old}, it was first predicted that 
weakly bound \textit{macrodimers} could be formed from the induced 
Van der Waals interactions of two Rydberg atoms. However, we have shown more 
recently~\cite{Samboy,Samboy-JpB} that larger, more stable
dimers can be formed \textit{via} the strong $\ell$-mixing of various 
Rydberg states. Recent measurements~\cite{shaffer-NPHYS} have shown
signatures of such macrodimers using an ultracold sample of cesium Rydberg atoms.

More recently, the focus of study has moved toward few-body interactions, such as between atom-diatom 
interactions~\cite{Byrd-Li3,Atom-diatom,Lewandowski11} and
diatom-diatom interactions~\cite{Byrd-NaK,Byrd-KRb}. Coinciding with this shift,
there have been proposals~\cite{Sadeghpour11,Sadeghpour10,Rost-poly,Jovica-trimers} 
for many-body long-range interactions involving Rydberg atoms. However,
these works focus on the interactions between one
Rydberg atom and ground state atoms or molecules. In this paper, 
we describe the long-range interactions between three Rydberg atoms arranged along 
a common axis and provide calculations, which predict the existence of bound trimer states. 
Here, we also present the lowest vibrational energies of these bound states, 
calculated \textit{via} the oscillation eigenmodes of the bound system.
%%%%%%%%%%%%%%%%%%%%%%%%%%%%%%%%%%%%%%%%%%%%%%%%%%%%%%%%%%%%%%%
\section{Three Body Interactions}
\label{sec:3Bod}
In~\cite{Samboy} and~\cite{Samboy-JpB}, we predicted 
the existence of long-range rubidium Rydberg dimers by analyzing potential energy curves
corresponding to the interaction energies between the two Rydberg atoms. In these works, 
we diagonalized an interaction Hamiltonian consisting of the long-range Rydberg-Rydberg 
interaction energy and atomic fine structure in the Hund's case (c) basis set. 
Each molecular state in the basis was symmetrized with respect to the $D_{\infty h}$ symmetry 
of the homonuclear dimer.

\begin{figure}[h!]
	\centering
		\includegraphics[width=3.75in]{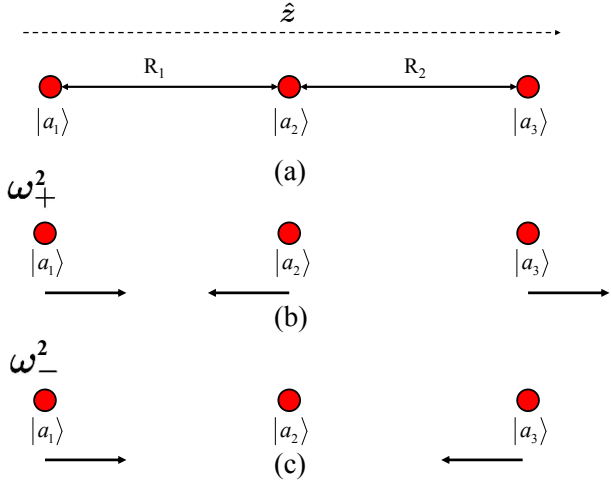}
	\caption{(Color online) (a) Three Rydberg atoms lie along a common $z$-axis. The distance 
	         between atoms 1 and 2 is represented by $R_1$ and the distance between atoms 2
	         and 3 is represented by $R_2$. Each atom is in state $\ket{a_i}$, defined in the text.
	         Each bound \textit{macrotrimer} has two eigenmodes of oscillation:
           (b) $\omega_{+}^2$ and (c) $\omega_{-}^2$ . (See text)
	        }
	\label{fig:Trimer}
\end{figure}

In general, adding a third atom to the interaction picture will change the physical 
symmetry of the system. However, to simplify our calculations, we assume that three 
identical Rydberg atoms lie along a common ($z$-) axis 
(see Fig.~\ref{fig:Trimer}(a)). 
This preserves the $D_{\infty h}$ 
symmetry and permits the use of much of the two-body physics
on the three-body system. We analyzed this symmetry state for three Rb $58s$ atoms and three Rb
$58p$ atoms and found that the $58p$ case exhibited examples of three-dimensional wells, 
indicating that the three atoms are bound in a linear chain.
%%%%%%%%%%%%%%%%%%%%%%%%%%%%%%%%%%%%%
%%%%%%%%%%%%%%%%%%%%%%%%%%%%%%%%%%%%%
\subsection{Basis States}
\label{subs:basis}
Obtaining properly symmetrized basis functions for the three-atom case is very similar to that 
of the two-atom system (see~\cite{Jovica}), but much more technically demanding. We construct
the molecular wave functions from three free Rydberg atoms in respective states
$\ket{a_1}\equiv\ket{n_1,\ell_1,j_1,m_{j_1}}$, $\ket{a_2}\equiv\ket{n_2,\ell_2,j_2,m_{j_2}}$,
and $\ket{a_3}\equiv\ket{a_3,\ell_3,j_3,m_{j_3}}$, where $n_i$ is the principal quantum 
number of atom $i$, $\ell_i$ is the orbital angular momentum of atom $i$, and $m_{j_i}$
is the projection of the total angular momentum $\vect{j}_i=\vect{\ell}_i+\vect{s}_i$ of atom
$i$ onto a quantization axis (chosen in the $z$-direction). 
As in the two-atom case, we assume that the three Rydberg atoms 
interact \textit{via} long-range dipole-dipole and quadrupole-quadrupole couplings. %which
%results in the mixing of different electronic states. 
Here, long-range indicates that the distance between each Rydberg atom is greater than the 
Le-Roy radius~\cite{LeRoy}:
\begin{equation}
\label{eq:LeRoy}
R_{LR}=2\,[\element{n_1\ell_1}{r^2}{n_1\ell_1}^{1/2}+
\element{n_2\ell_2}{r^2}{n_2\ell_2}^{1/2}]\,\,,
\end{equation}
such that there is no overlapping of the electron clouds.
The properly symmetrized long-range three-atom wavefunctions take the form:
\begin{align}
\label{eq:3Bodywf}
\ket{a_1;a_2;a_3}= \phantom{BLAHBLAH}& \nonumber\\
\dfrac{1}{\sqrt{6}}\left[\left(\ket{a_1}_1\ket{a_2}_2\ket{a_3}_3 \right.\right. &+ \ket{a_2}_1\ket{a_3}_2\ket{a_1}_3 \nonumber\\
    &+ \left.\ket{a_3}_1\ket{a_1}_2\ket{a_2}_3\right) \\
-P\left(\ket{a_1}_1\ket{a_3}_2\ket{a_2}_3 \right. &+ \ket{a_2}_1\ket{a_1}_2\ket{a_3}_3 \nonumber\\
   &+ \left.\left.\ket{a_3}_1\ket{a_2}_2\ket{a_1}_3\right)\right]\nonumber\,\,,
\end{align}
%\begin{equation}
%\label{eq:3Bodywf}
%\ket{a_1;a_2;a_3}=\dfrac{1}{\sqrt{6}}\,{\displaystyle\sum_{i,j,k}}
%\ket{a_i}_1\ket{a_j}_2\ket{a_k}_3\,\epsilon_{ijk}\,P^{(1-\epsilon_{ijk})/2},
%\end{equation}
%where the sums over $i$, $j$, and $k$ run from 1 to 3, $\epsilon_{ijk}$ is the levi-cevita symbol, 
where $P=p(-1)^{\ell_1+\ell_2+\ell_3}$, with $p=+1(-1)$ for gerade (ungerade) molecular states.

%We diagonalize an interaction Hamiltonian in a 
The basis
set consists of the Rydberg molecular level being probed (\textit{e.g.} $58p+58p+58p$), as well
as all nearby asymptotes with significant coupling to this level and to each other. All of these
states are properly symmetrized \textit{via} Equation~\eqref{eq:3Bodywf}
according to their molecular symmetry $\Omega=m_{j_1}+m_{j_2}+m_{j_3}$. In this paper, we 
consider the $\Omega = 1/2$ symmetry.
%%%%%%%%%%%%%%%%%%%%%%%%%%%%%%%%%%%%%%%%%%%%%%%%%%%%%%%%%%%%%%%%%%
\subsection{Interaction Hamiltonian}
\label{subs:interactions}
As in the two-atom case, we construct the interaction picture for the three-atom system by diagonalizing
an interaction Hamiltonian in the properly symmetrized basis described in the previous subsection.
The Hamiltonian consists
of a three-body long-range Rydberg interaction energy and atomic fine structure,
\textit{i.e.} $H_{\rm{int}}=V_{3-\rm{body}}+H_{fs}$.
Using the wave functions defined by Equation~\eqref{eq:3Bodywf}, we write the matrix elements 
of the Hamiltonian as the sums of multiple interactions. Each matrix element is defined as:
\begin{align}
\label{eq:3BVint}
\element{a_1;a_2;&a_3}{V_{3-\rm{body}}}{b_1;b_2;b_3} =\nonumber\\ 
&\frac{1}{6}\displaystyle\sum_{\substack{i,j,k\\i',j',k'}}\element{a^{(1)}_i a^{(2)}_j a^{(3)}_k}
{V_{3-\rm{body}}}{b^{(1)}_{i'}b^{(2)}_{j'}b^{(3)}_{k'}}\\ &\times\left(\Theta_C+P_a\Theta_A\right)
\left(\Theta_C+P_b\Theta_A\right)\nonumber \,\,,
\end{align}
where each summation index is over the total number of atoms, \textit{i.e.} from 1 to 3,
$P$ is as before, we have defined
\begin{equation*}
\begin{array}{ccc}
 & \Theta_C= \left\{\begin{array}{rl}
                  -1 & \mbox{for cyclic permutations}\\
                  0  & \mbox{for anti-cyclic permutations}\\
                  \end{array} \right. & \\
                & &\\
                &  \mbox{and} & \\
                & & \\
&  \Theta_A= \left\{\begin{array}{rl}
                  0 & \mbox{for cyclic permutations}\\
                  -1  & \mbox{for anti-cyclic permutations}\\
                  \end{array} \right. &
\end{array}
\end{equation*}
and we have defined $\ket{a^{(1)}_ia^{(2)}_ja^{(3)}_k}\equiv\ket{a_i}_1\ket{a_j}_2\ket{a_k}_3$, etc.
In the case that $\ket{a_1;a_2;a_3}=\ket{b_1;b_2;b_3}$ 
(\textit{i.e.} along the diagonal of the matrix), the matrix element is given by:
\begin{align}
\element{a_1;a_2;a_3&}{H_{\rm int}}{a_1;a_2;a_3}= \nonumber \\
&\element{a_1;a_2;a_3}{V_{3-\rm{body}}}{a_1;a_2;a_3}+E_{123}\,\,,
\end{align}
with $E_{123}=E_1+E_2+E_3$, where 
%$E_k=-\,\dfrac{1}{2\,(n_k-\delta_{\ell_k})^2}$ 
$E_k=-\frac{1}{2}(n_k-\delta_{\ell_k})^{-2}$
are the atomic Rydberg energies with respective quantum defects $\delta_{\ell_k}$.

The long-range assumption assures that these are three free atoms interacting \textit{via}
long-range two-body potentials. That is, the transition element
$\element{a^{(1)}_i a^{(2)}_j a^{(3)}_k}{V_{3-\rm{body}}}
{b^{(1)}_{i'}b^{(2)}_{j'}b^{(3)}_{k'}}$ given in Equation~\eqref{eq:3BVint}
is defined as a sum of two-body interactions:
\begin{align}
\element{a^{(1)}_i a^{(2)}_j a^{(3)}_k}
{&V_{3-\rm{body}}}{b^{(1)}_{i'}b^{(2)}_{j'}b^{(3)}_{k'}} =\nonumber\\
&\phantom{+}\;\,\element{a^{(1)}_i a^{(2)}_j}{V_L(R_{12})}
{b^{(1)}_{i'} b^{(2)}_{j'}}\nonumber\\
&+\element{a^{(2)}_ja^{(3)}_k}{V_L(R_{23})}{b^{(2)}_{j'} b^{(3)}_{k'}}\nonumber\\
&+\element{a^{(1)}_ia^{(3)}_k}{V_L(R_{13})}{b^{(1)}_{i'} b^{(3)}_{k'}}\,\,,
\end{align}
where $R_{\alpha\beta}$ is the nuclear separation between atoms $\alpha$ and $\beta$, and
$L=1(2)$ for dipolar (quadrupolar) interactions.
Since we are assuming that the three atoms lie along a common axis, each two-body 
interaction term
$\element{a^{(\alpha)}_ia^{(\beta)}_j}{V_L(R_{\alpha\beta})}
{b^{(\alpha)}_{i'}b^{(\beta)}_{j'}}$
defines the long-range transition element between 
the two respective Rydberg atoms. Each transition element is given by~\cite{Jovica,Samboy-JpB}:
\begin{align}
\element{1&2}{V_L(R)}{34}=\nonumber\\
&(-1)^{L-1-m_{j_{\rm tot}}+j_{\rm tot}}
\sqrt{\hat{\ell}_1\hat{\ell}_2\hat{\ell}_3\hat{\ell}_4\hat{j}_1\hat{j}_2\hat{j}_3\hat{j}_4}
\;\dfrac{\mathcal{R}_{13}^L\;\mathcal{R}_{24}^L}{R^{2L+1}}\nonumber\\
&\times\ThreeJ{\ell_1}{L}{\ell_3}{0}{0}{0}\ThreeJ{\ell_2}{L}{\ell_4}{0}{0}{0}\nonumber\\
&\times\SixJ{j_1}{L}{j_3}{\ell_3}{\frac{1}{2}}{\ell_1}
\SixJ{j_2}{L}{j_4}{\ell_4}{\frac{1}{2}}{\ell_2}\nonumber\\
&\times{\displaystyle\sum_{m=-L}^{L}}B_{2L}^{L+m}\ThreeJ{j_1}{L}{j_3}{-m_{j_1}}{m}{m_{j_3}}\nonumber\\
&\times\ThreeJ{j_2}{L}{j_4}{-m_{j_2}}{-m}{m_{j_4}}\,\,,
\end{align}
where $j_{\rm tot}=j_1+j_2+j_3+j_4$, $m_{j_{\rm tot}}=m_{j_1}+m_{j_2}+m_{j_3}+m_{j_4}$, 
$\hat{\ell}_i=2\ell_i+1$, $\hat{j}_i=2j_i+1$,
and $\mathcal{R}_{ij}^L=\element{i}{r^L}{j}$ is the radial matrix element. 
%%%%%%%%%%%%%%%%%%%%%%%%%%%%%%%%%%%%%%%%%%%%%%%%%%%%%%%%%%%%%%%%%%%%%%%%%%%
\section{Potential Energy Surfaces}
\label{sec:PES}
\subsection{General Cases}
\label{subs:General}
We diagonalize the three-body Hamiltonian at successive values of $R_1$ and $R_2$, 
resulting in a series of potential energy surfaces (PES), where
each surface corresponds to a different molecular asymptote in the basis. 
In each of the plots shown, $R_1$ represents
the distance between atom 1 and atom 2, $R_2$ represents the distance between
atom 2 and atom 3, both in $a_0$ (see Fig.~\ref{fig:Trimer}(a)) 
and the energy is measured in GHz. The color scheme for the energy values is given in the 
scales to the right of each plot.

\begin{figure}[h]
	\centering
		\includegraphics[width=3.5in]{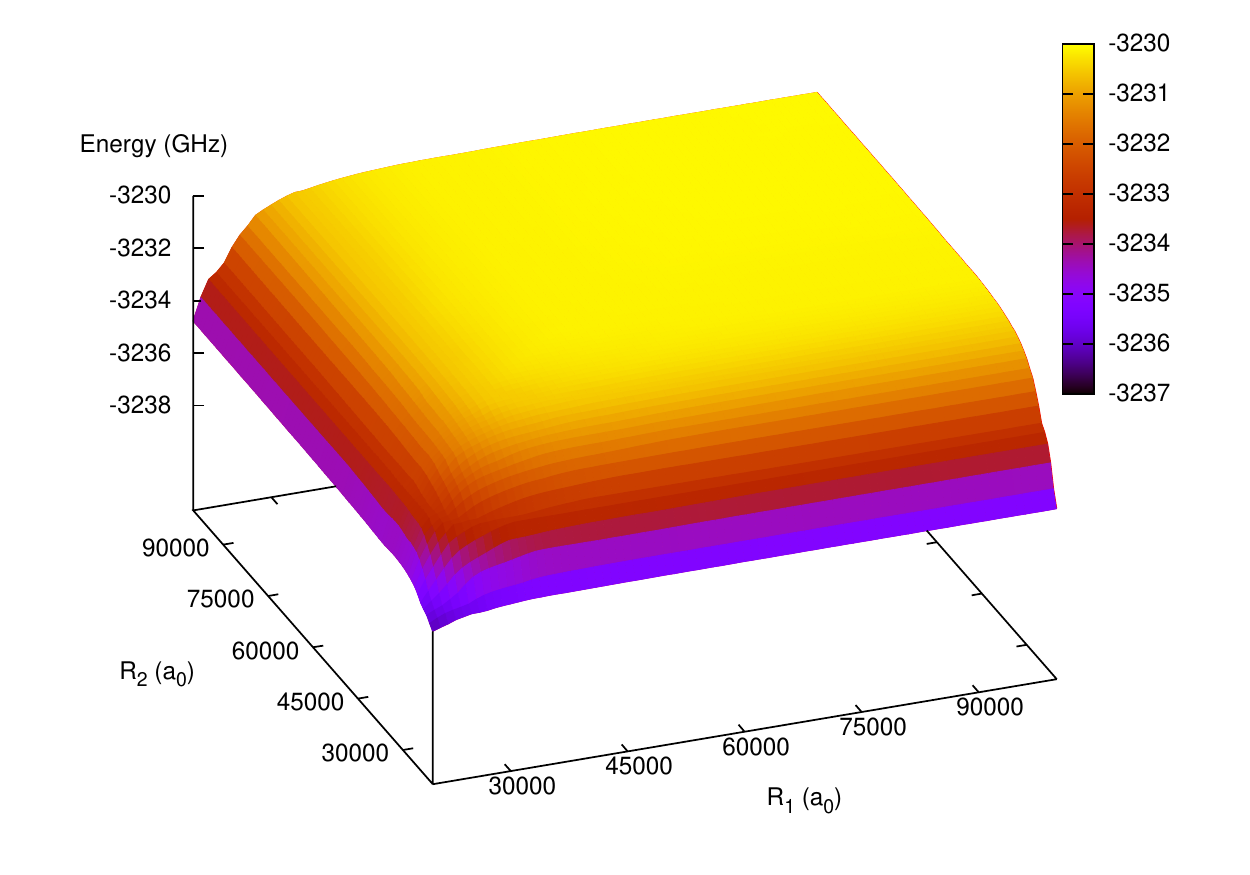}
	\caption{(Color online) Potential energy surface (PES) correlated to the
	         $\ket{56p\frac{1}{2},\frac{1}{2};58p\frac{3}{2},-\frac{1}{2};60p\frac{3}{2},\frac{1}{2}}$
	         asymptotic state. This surface is analogous to a repsulsive potential curve for the
	         two-body case: As the distance of either the first or last atom in the 
	         linear chain is increased, the two local atoms remain repulsed. The color scheme
	         denotes the energy values given in GHz, with the scale presented to the right of
	         the plot.}
		\label{fig:repulse}
\end{figure}

\begin{figure}[!h]
	\centering
		\includegraphics[width=3.5in]{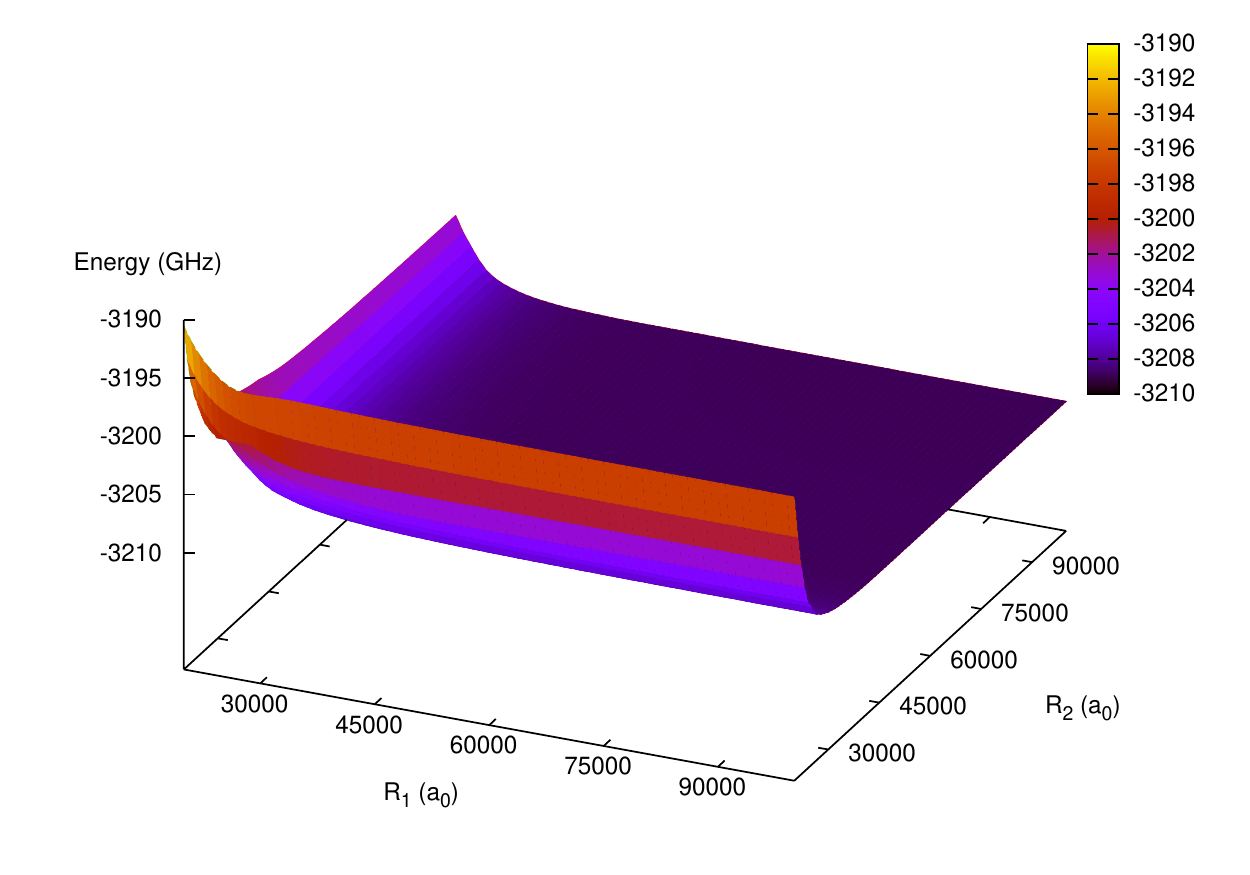}
	\caption{(Color online) Potential energy surface (PES) correlated to the
	         $\ket{58s\frac{1}{2},\frac{1}{2};59s\frac{1}{2},-\frac{1}{2};57d\frac{5}{2},\frac{1}{2}}$
	         asymptotic state. This surface is analogous to an attractive potential curve for the
	         two-body case: As the distance of either the first or last atom in the 
	         linear chain is increased, the two local atoms remain attracted. The color scheme
	         denotes the energy values given in GHz, with the scale presented to the right of
	         the plot.}
		\label{fig:attract}
\end{figure}

As a result of the large 
$\ell$-mixing that occurs between the Rydberg atoms, these surfaces have interesting topographies.
For example, Figures~\ref{fig:repulse} and~\ref{fig:attract} illustrate potential surfaces 
analogous to two-dimensional repulsive and attractive curves, respectively. The repulsive PES shown in 
Fig.~\ref{fig:repulse} corresponds to the 
$\ket{56p\frac{1}{2},\frac{1}{2};58p\frac{3}{2},-\frac{1}{2};60p\frac{3}{2},\frac{1}{2}}$
state, while the attractive PES shown in 
Fig.~\ref{fig:attract} corresponds to the 
$\ket{58s\frac{1}{2},\frac{1}{2};59s\frac{1}{2},-\frac{1}{2};57d\frac{5}{2},\frac{1}{2}}$ 
state. We see that in both cases, 
the distance of the third atom has very
little effect on the other two atoms: as either $R_1$ or $R_2$ is increased (while keeping the other 
distance fixed), the two stationary atoms consistently demonstrate an attractive/repulsive
behavior.
%%%%%%%%%%%%%%%%%%%%%%%%%%%%%%%%%%%%%%%%%%%%
%%%%%%%%%%%%%%%%%%%%%%%%%%%%%%%%%%%%%%%%%%%%
%%%%%%%%%%%%%%%%%%%%%%%%%%%%%%%%%%%%%%%%%%%%

\begin{figure}[h]
	\centering
		\includegraphics[width=3.5in]{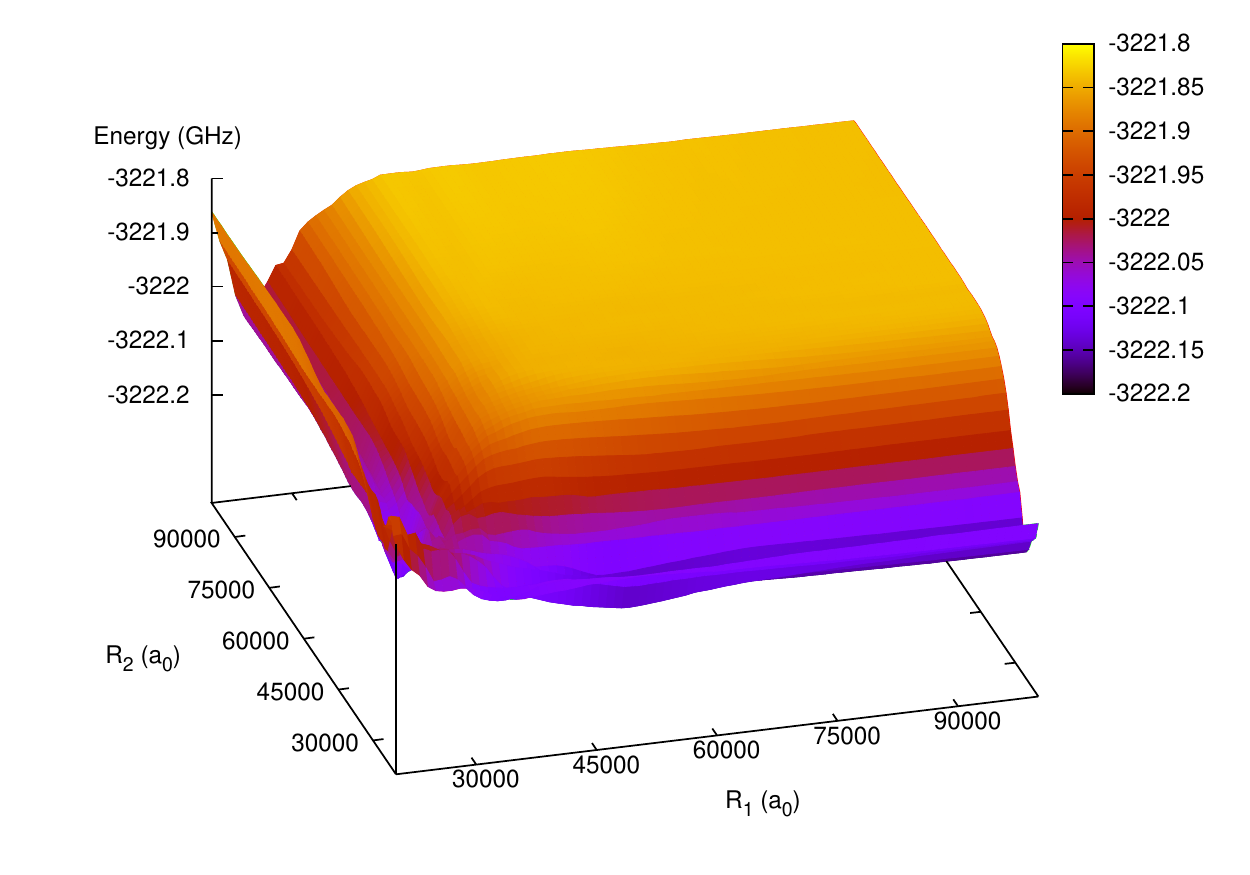}
	\caption{(Color online) Potential energy surface (PES) correlated to the 
	         $\ket{59s\frac{1}{2},-\frac{1}{2};55d\frac{3}{2},\frac{3}{2};58d\frac{3}{2},-\frac{1}{2}}$
	         asymptotic state. We note the ``ridge'' lying along the $R_2$ axis, which indicates that
	         atom 1 and atom 2 are bound (see text). The color scheme
	         denotes the energy values given in GHz, with the scale presented to the right of
	         the plot.}
	\label{fig:ridge}
\end{figure}

Figure~\ref{fig:ridge} illustrates another type of surface, in which there is a significant
``ridge'' running along one of the axes (in this case along the $R_2$ axis). 
Such a ridge indicates that the two local atoms (\textit{e.g.} atom 1 and atom 2) form 
a bonded pair, existing even as atom 3 is moved away. This particular surface corresponds
to the 
$\ket{59s\frac{1}{2},-\frac{1}{2};55d\frac{3}{2},\frac{3}{2};58d\frac{3}{2},-\frac{1}{2}}$
asymptotic state.
%%%%%%%%%%%%%%%%%%%%%%%%%%%%%%%%%%%%%%%%%%%
%%%%%%%%%%%%%%%%%%%%%%%%%%%%%%%%%%%%%%%%%%%
\subsection{Potential Wells}
\label{subs:PotWells}
\begin{figure}[!h]
	\centering
		\includegraphics[width=3.5in]{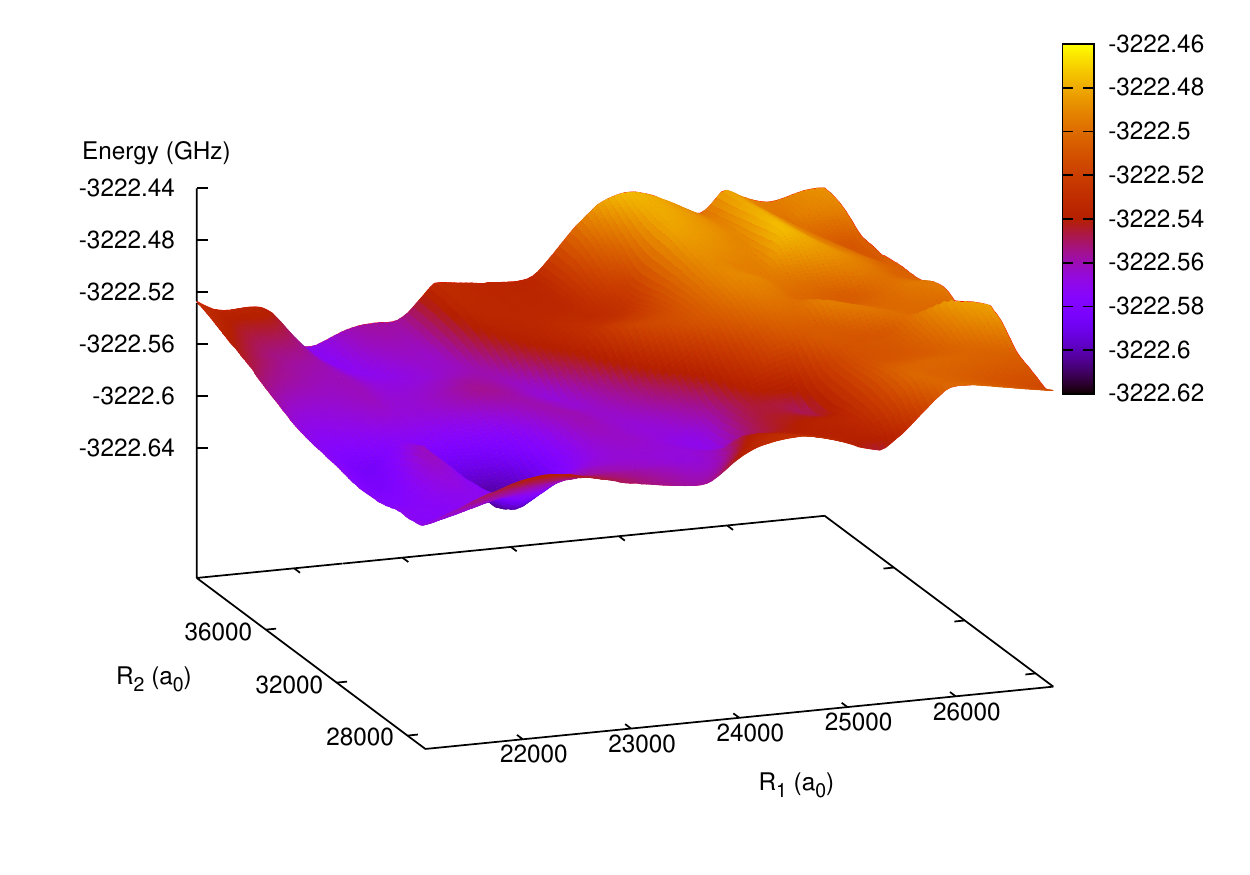}
		\includegraphics[width=3.25in]{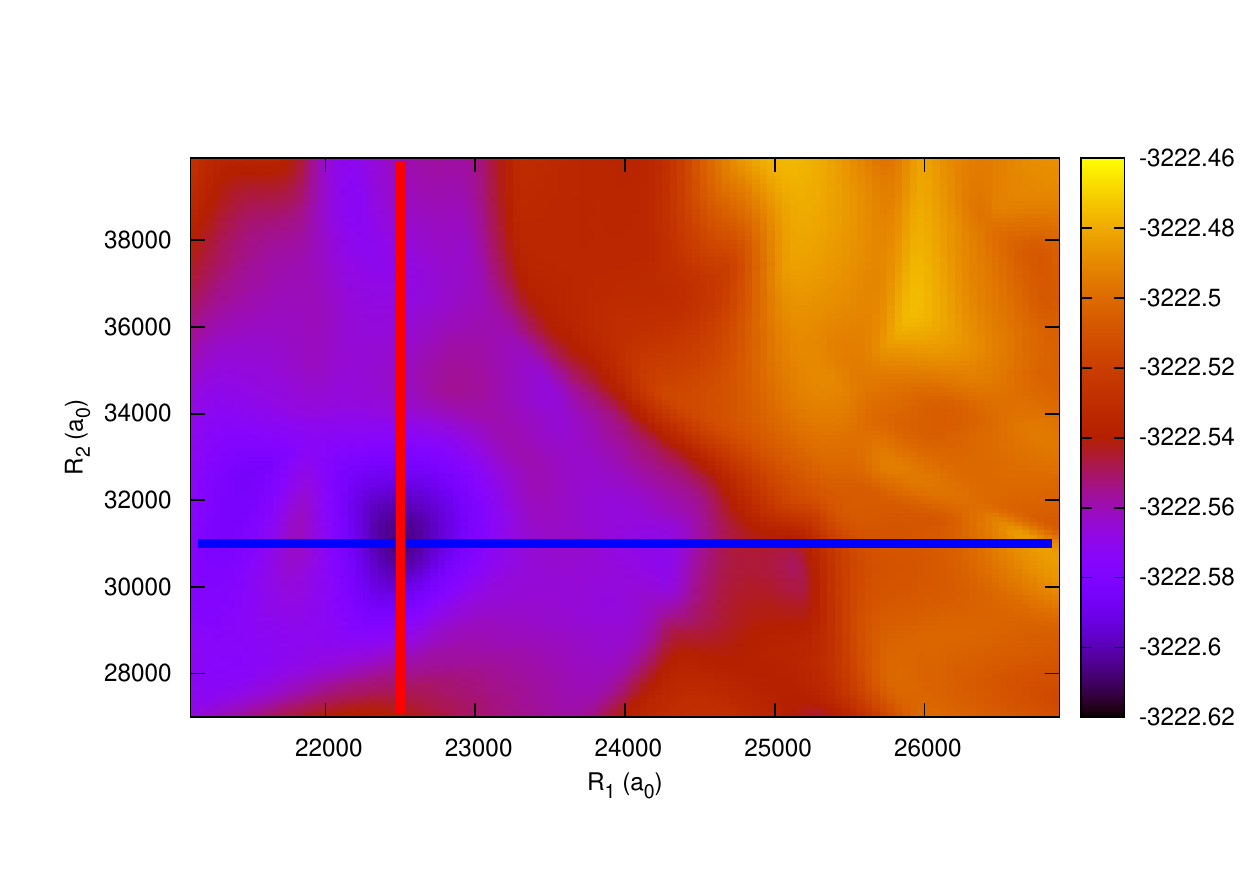}
	\caption{(Color online) Potential energy surface (top) and two-dimensional 
	         projection (bottom) correlated to the 
	         $\ket{58p\frac{1}{2}\frac{1}{2};58p\frac{1}{2},\frac{1}{2};58p\frac{1}{2},-\frac{1}{2}}$
	         asymptotic molecular state. The main
	         feature of these plots is the three-dimensional well, centered at 
	         $R_{1e}=22$~500~$a_0$ and $R_{2e}=31$~000~$a_0$, indicating that the three
	         Rydberg atoms are bound together in a linear chain (see text). The red (vertical)
	         and blue (horizontal) lines imposed on the bottom plot are centered at the well minimum and 
	         indicate the cross-sections for which quadratic fits are performed 
	         (see Figure~\ref{fig:Harmonic}). The color scheme
	         denotes the energy values given in GHz, with the scale presented to the right of
	         each plot. 
	         }
	\label{fig:TrimerWell}
\end{figure}

Although the surfaces highlighted in the previous section are interesting, ultimately we seek
surfaces that illustrate potential wells, as these indicate bound three-atom systems. Upon 
separately investigating three excited $58s$ rubidium atoms and three excited $58p$ rubidium atoms,
we found that in the case of the three excited $58p$ atoms, surface plots corresponding to various
$58p+58p+58p$ asymptotes illustrated such (three-dimensional) potential wells. 
Specifically, wells were determined for the following states:\\
\\
$\ket{1}\equiv\ket{58p\frac{1}{2},\frac{1}{2};58p\frac{1}{2},\frac{1}{2};58p\frac{1}{2},-\frac{1}{2}}$\\
$\ket{2}\equiv\ket{58p\frac{1}{2},-\frac{1}{2};58p\frac{1}{2},-\frac{1}{2};58p\frac{3}{2},\frac{3}{2}}$\\
$\ket{3}\equiv\ket{58p\frac{1}{2},\frac{1}{2};58p\frac{3}{2},\frac{1}{2};58p\frac{3}{2},-\frac{1}{2}}$\\
$\ket{4}\equiv\ket{58p\frac{1}{2},-\frac{1}{2};58p\frac{3}{2},-\frac{1}{2};58p\frac{3}{2},\frac{3}{2}}$\\
$\ket{5}\equiv\ket{58p\frac{3}{2},-\frac{1}{2};58p\frac{3}{2},-\frac{1}{2};58p\frac{3}{2},\frac{3}{2}}$\,\,.\\
\\
As an example, in Figure~\ref{fig:TrimerWell} we show the PES and the corresponding 
two-dimensional projection correlated to the 
$\ket{1}$ state, which demonstrates a potential well
approximately 50-100 MHz deep.
Due to the large equilibrium separations, \textit{e.g.}
$R_{1e}\sim 22$~500~$a_0$ and $R_{2e}\sim 31$~000~$a_0$,
we label these bound states \textit{macrotrimers}.
%%%%%%%%%%%%%%%%%%%%%%%%%%%%%%%%%%%%%%
%%%%%%%%%%%%%%%%%%%%%%%%%%%%%%%%%%%%%%%
%As shown in Figure~\ref{fig:Trimer}, we model the linear trimer as a classical
%mass-spring system, where each
%outer mass is connected to the inner mass \textit{via} a spring with a unique spring constant $k$,
%\textit{i.e.}, $k_1\ne k_2$. Since these are identical Rydberg atoms, the masses are all
%the same. Ignoring translational motion, the eigenvalues $\omega^2$ of
%the oscillations for such a classical system are found to be
%\begin{equation}
%\omega^2_{(\pm)} = \dfrac{(k_1+k_2)\pm\sqrt{k_1^2-k_1 k_2 + k_2^2}}{m}\,\,,
%\label{eq:modes}
%\end{equation}
%where $m$ is the mass of a single Rydberg atom. 

For the three-atom configuration shown in Figure~\ref{fig:Trimer}(a), the (non-zero) oscillation modes can be 
calculated \textit{via}:
\begin{equation}
\omega^2_{(\pm)} = \dfrac{(k_1+k_2)\pm\sqrt{k_1^2-k_1 k_2 + k_2^2}}{m}\,\,,
\label{eq:modes}
\end{equation}
where $k_i$ are ``effective spring constants'' that need to be calculated and $m$ is the mass of a single rubidium atom.
In Figure~\ref{fig:Trimer}(b) and (c), we show the physical description of each eigenmode.
Panel (b) corresponds to the $\omega^2_+$ eigenvalue and shows that the two outer Rydberg atoms 
vibrate in the same direction, opposite to the inner atom's direction of motion. 
Panel (c) corresponds to the $\omega^2_-$ eigenvalue and shows that the inner Rydberg atom is
stationary while the outer two atoms oscillate in opposing directions. 

%%%%%%%%%%%%%%%%%%%%%%%%%%%%%%%%%%%%%%%%%%%%%%%%%%%%%%%%%%%%%%%
\begin{table}[h]
\caption{Characteristics of the polynomial fitting procedures for each trimer state $\ket{i}$ (see text), 
         assumed to be quadratic.
         A two-dimensional cross-section was taken
         at the minimum of each well for both the $R_1$ and $R_2$ axes, where the well is
         centered ($R_{1e}$, $R_{2e}$). The 
         $k_{\rm eff}$-values (in N/m) 
         correspond to the numeric fitting along each respective axis. The well depth indicates the 
         potential energy range for which the quadratic assumption is valid and for which the given $r^2$ 
         goodness-of-fit values are appropriate.
         }
\begin{tabular}{c|c|c|c|c|c}
\hline\hline
State & Axis & $R_e(\times 10^3 a_0$) & $k_{\rm eff}$ (N/m) & $r^2$-value & Depth (MHz)\\
\hline
\multirow{2}{*}{$\ket{1}$} & $R_1$ & 22.50 & $8.37\times 10^{-11}$ & 0.9879 & 17.50 \\ \cline{2-6}
                           & $R_2$ & 31.00 & $4.40\times 10^{-12}$ & 0.9926 & 16.60 \\
\hline
\multirow{2}{*}{$\ket{2}$} & $R_1$ & 23.15 & $3.85\times 10^{-11}$ & 0.9861 & 16.10 \\ \cline{2-6}
                           & $R_2$ & 33.10 & $1.07\times 10^{-12}$ & 0.9781 & 77.50 \\
\hline                           
\multirow{2}{*}{$\ket{3}$} & $R_1$ & 22.80 & $2.55\times 10^{-11}$ & 0.9908 & 37.70 \\ \cline{2-6}
                           & $R_2$ & 34.60 & $3.02\times 10^{-12}$ & 0.9892 & 21.30 \\
\hline
\multirow{2}{*}{$\ket{4}$} & $R_1$ & 22.85 & $4.63\times 10^{-11}$ & 0.9985 & 29.30 \\ \cline{2-6}
                           & $R_2$ & 31.50 & $5.54\times 10^{-12}$ & 0.9943 & \z5.80 \\
\hline                           
\multirow{2}{*}{$\ket{5}$} & $R_1$ & 22.00 & $6.87\times 10^{-11}$ & 0.9982 & 16.24 \\ \cline{2-6}
                           & $R_2$ & 31.70 & $1.72\times 10^{-12}$ & 0.9992 & 11.94 \\                           
\hline\hline                           
\end{tabular}
\label{tab:fitparams}
\end{table}
%%%%%%%%%%%%%%%%%%%%%%%%%%%%%%%%%%%%%%%%%%%%%%%%%%%%%%%%%%%%%%%

\begin{figure}[!h]
	\centering
		\includegraphics[width=3.5in]{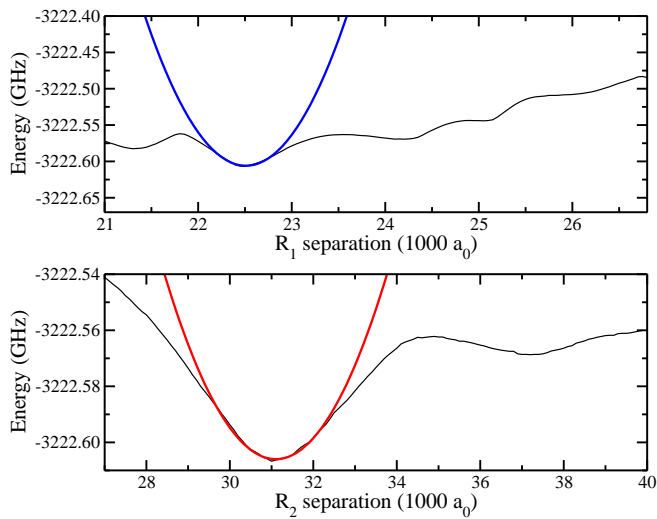}
	\caption{(Color online) Two-dimensional cross sections of the three dimensional surface shown in 
	          Figure~\ref{fig:TrimerWell} taken along the $R_1$ (top) and $R_2$ (bottom) axes at the 
	          well's minimum. Each cross-section is fitted with a harmonic potential centered at the 
	          minimum. The results of these fits are summarized in Table~\ref{tab:fitparams}.
	         }
	\label{fig:Harmonic}
\end{figure}

To calculate the $k_i$ for Equation~\eqref{eq:modes}, we perform polynomial fits to two-dimensional 
cross sections of the potential wells along the $R_1$ and $R_2$ axes at each well's minima
(see Fig.~\ref{fig:Harmonic}). The deepest 
portions of each well can be modeled as a simple harmonic oscillator, and it is easily shown
that the desired $k$-value equals the second derivative of these quadratic fits (with respect to
the nuclear separation in that particular direction), \textit{i.e.}
\begin{equation}
\left.k_i = \frac{d^2V}{dR_i^2}\right|_{R_{ie}}\,\,,
\end{equation}
where $R_{ie}$ is the equilibrium separation along axis $R_i$.
Table~\ref{tab:fitparams} summarizes the results of the polynomial fitting 
including the $k_{\rm eff}$-values, the goodness-of-fits ($r^2$ values) 
of the harmonic oscillator potentials, 
and the potential energy range for which the quadratic assumption is valid.
%%%%%%%%%%%%%%%%%%%%%%%%%%%%%%%%%%%%%%%%%%%%%%%%%%%%%%%%%%%%%%%%%
%%%%%%%%%%%%%%%%%%%%%%%%%%%%%%%%%%%%%%%%%%%%%%%%%%%%%%%%%%%%%%%%%

Based on these properties, we use Equation~\eqref{eq:modes} and the familiar $E=\hbar\omega(v+1/2)$ 
to find the oscillation energies in the deepest portions of the highlighted wells.
Table~\ref{tab:TriEnergies}
lists the vibrational energies of the first few bound states for each modal frequency,
$\omega_{+}$ and $\omega_{-}$. We see that in each case, the energies defined by the $\omega_+$
frequency illustrate spacings of about 3-6 MHz, which are separated enough
to be detected through spectroscopic means. The MHz energy values correspond to $\mu$s
oscillation periods, which are rapid enough to allow for several oscillations during the 
lifetimes of these Rydberg atoms (roughly 500~$\mu$s for $n\sim 60$~\cite{RydLife}). The
energies corresponding to the $\omega_-$ frequency are more closely spaced 
and demonstrate oscillation periods that are slower, but should still be able to be
detected experimentally.
%%%%%%%%%%%%%%%%%%%%%%%%%%%%%%%%%%%%
\begin{table}[!h]
	\caption{Lowest vibrational levels and corresponding bound state energies for 
	         both oscillation frequencies of each trimer state $\ket{i}$ (see text). 
	         Energy$_{\pm}$ represents the bound energies
	         associated with $\omega_{\pm}$ and is measured from the bottom of the 
	         potential well. 
	         }
	\centering
		\begin{tabular}{c|c|c|c}
		\hline\hline
			State & $v$ & Energy$_{+}$ (MHz) & Energy$_{-}$ (MHz)\\
			\hline
\multirow{6}{*} {$\ket{1}$} & 0 &  \z2.71  & \z0.62 \\
			                      & 1 &  \z8.13  & \z1.86 \\
			                      & 2 &   13.55  & \z3.11 \\
			                      & 3 &  $-$     & \z4.34 \\
			               & $\vdots$ & $\vdots$ & $\vdots$\\
			                      & 12&  $-$     & 15.53 \\
\hline			 
\multirow{7}{*} {$\ket{2}$} & 0 &  \z1.84  & \z0.31 \\
			                      & 1 &  \z5.51  & \z0.92 \\
			                      & 2 &  \z9.18  & \z1.53 \\
			                      & 3 &   12.86  & \z2.14 \\
			                      & 4 &  $-$     & \z2.75 \\
			               & $\vdots$ & $\vdots$ & $\vdots$ \\
			                      &25 &  $-$     & 15.60 \\
\hline
\multirow{10}{*} {$\ket{3}$} & 0 &  \z1.50  & \z0.51 \\
			                      & 1 &  \z4.48  & \z1.54 \\
			                      & 2 &  \z7.48  & \z2.57 \\
			                      & 3 &   10.47  & \z3.60 \\
			                      & 4 &   13.47  & \z4.63 \\
			                      & 5 &   16.46  & \z5.66 \\
			                      & 6 &   19.45  & \z6.69 \\
			                      & 7 &  $-$     & \z7.72 \\   
			               & $\vdots$ & $\vdots$ &$\vdots$\\
			                     & 20 &    $-$   & 21.10 \\
\hline
\multirow{4}{*} {$\ket{4}$} & 0 &  \z2.02  & \z0.70 \\
			                      & 1 &   $-$    & \z2.09 \\
			                      & 2 &   $-$    & \z3.49 \\
			                      & 3 &   $-$    & \z4.88 \\
\hline
\multirow{5}{*} {$\ket{5}$} & 0 &  \z2.46  & \z0.39 \\
			                      & 1 &  \z7.37  & \z1.17 \\
			                      & 2 &   $-$    & \z1.94 \\
			               & $\vdots$ & $\vdots$ & $\vdots$ \\
			                      & 14 &   $-$    & 11.27 \\			                      
			 \hline\hline
		\end{tabular}
	\label{tab:TriEnergies}
\end{table}
%%%%%%%%%%%%%%%%%%%%%%%%%%%%%%%%%%
%$\omega_{(+)} = 25.725\times 10^6$ rad/s and $\omega_{(-)} = 1.24476\times10^6$ rad/s. 
\subsection{Other asymptotes}
\label{subs:others}
Of course, the potential wells described in section~\ref{subs:PotWells} are not the only wells that exist. During
the course of our analysis, we also found wells corresponding to various $\ket{59s55d58d}$ asymptotes; some examples of which
are presented below. Although in principle these wells can be evaluated in the same amount of detail as was done
for the $\ket{58p58p58p}$ wells in section~\ref{subs:PotWells}, we merely present visual evidence of their 
existence in this paper. 
Should these additional asymptotes lend themselves to specific experimental probing, then computing their respective energy 
levels would be of value. At this time, however, such evaluation is beyond the goals of this paper.
%%%%%%%%%%%%%%%%%%%%%%%%%%%%%%%%%%%%%%%%%%%%%%%%%%%%%%%%%%%%%%%
\begin{figure}[h]
	\centering
		\includegraphics[width=3.5in]{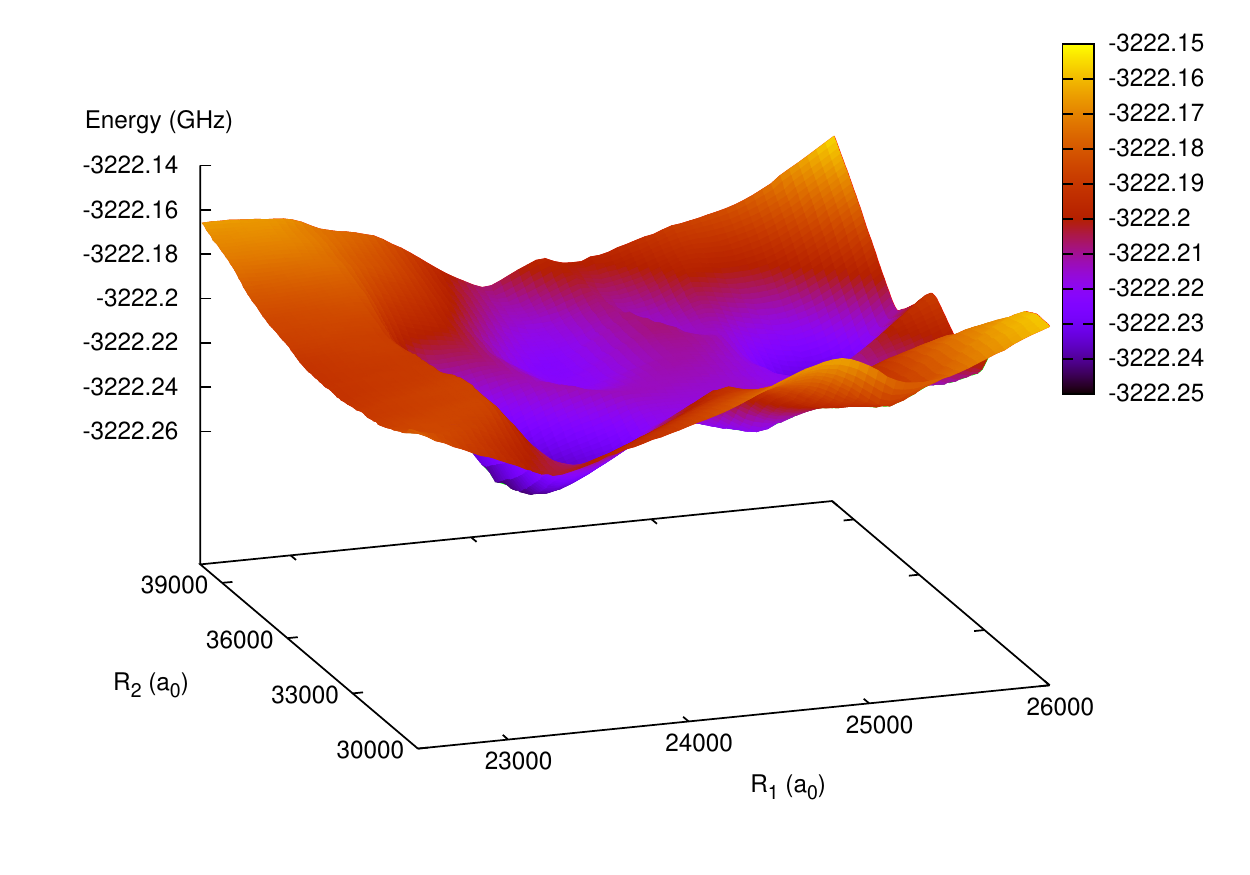}
	\caption{(Color online) Potential surface corresponding to the\\ 
	         $\ket{59s\frac{1}{2}-\frac{1}{2};55d\frac{3}{2}\frac{1}{2};58d\frac{3}{2}\frac{1}{2}}$ asymptotic state.
	         The color scheme
	         denotes the energy values given in GHz, with the scale presented to the right of
	         the plot.
	         This particular surface actually exhibits a few potential wells, the largest one being between 40-50 MHz deep.
	         }
	\label{fig:397}
\end{figure}
%%%%%%%%%%%%%%%%%%%%%%%%%
\begin{figure}[h]
	\centering
		\includegraphics[width=3.5in]{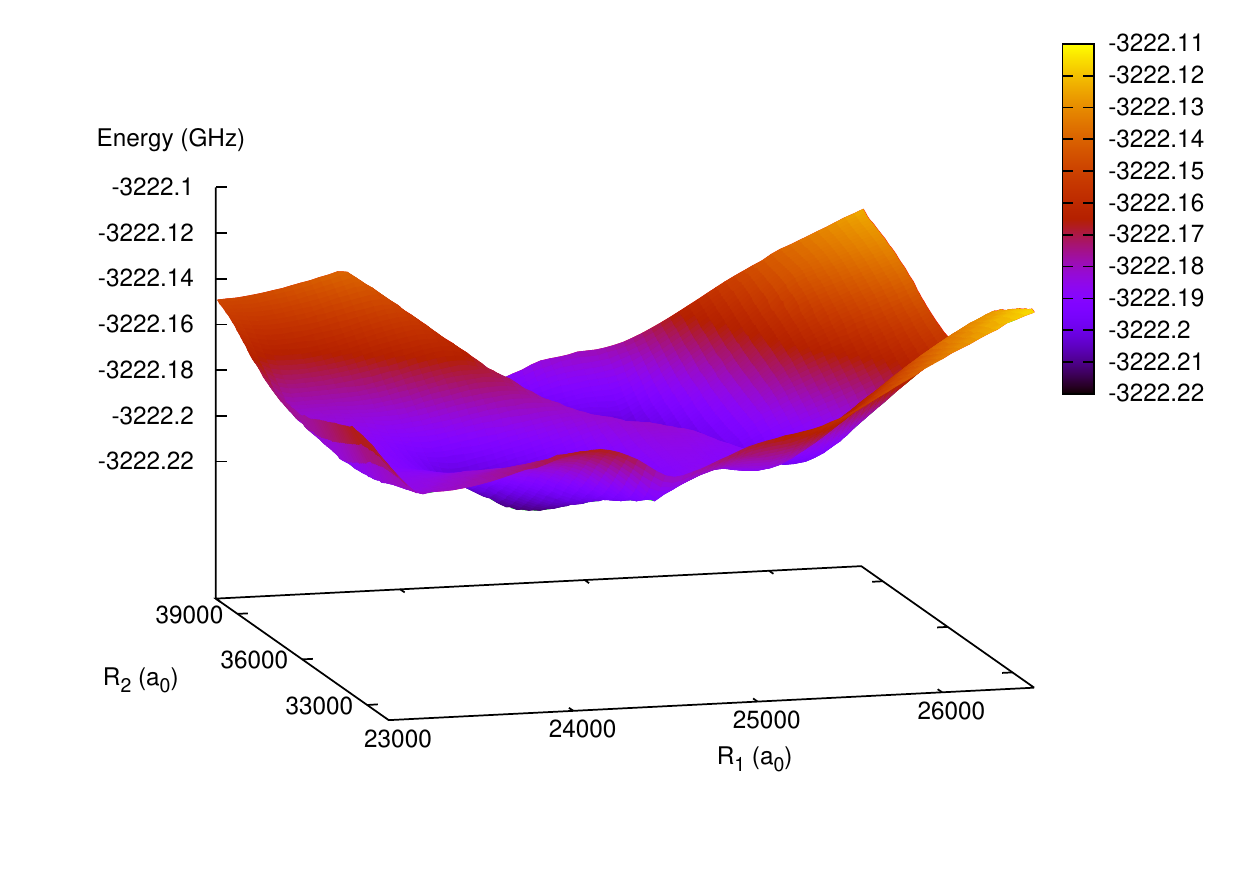}
	\caption{(Color online) Potential surface corresponding to the\\ 
	         $\ket{59s\frac{1}{2}-\frac{1}{2};55d\frac{3}{2}-\frac{1}{2};58d\frac{3}{2}\frac{3}{2}}$ asymptotic state.
	         The color scheme
	         denotes the energy values given in GHz, with the scale presented to the right of
	         the plot.
	         This particular surface exhibits two potential wells, each about 30-40 MHz deep.
	         }
	\label{fig:398}
\end{figure}
%%%%%%%%%%%%%%%%%%%%%%%%%%
\begin{figure}[h]
	\centering
		\includegraphics[width=3.5in]{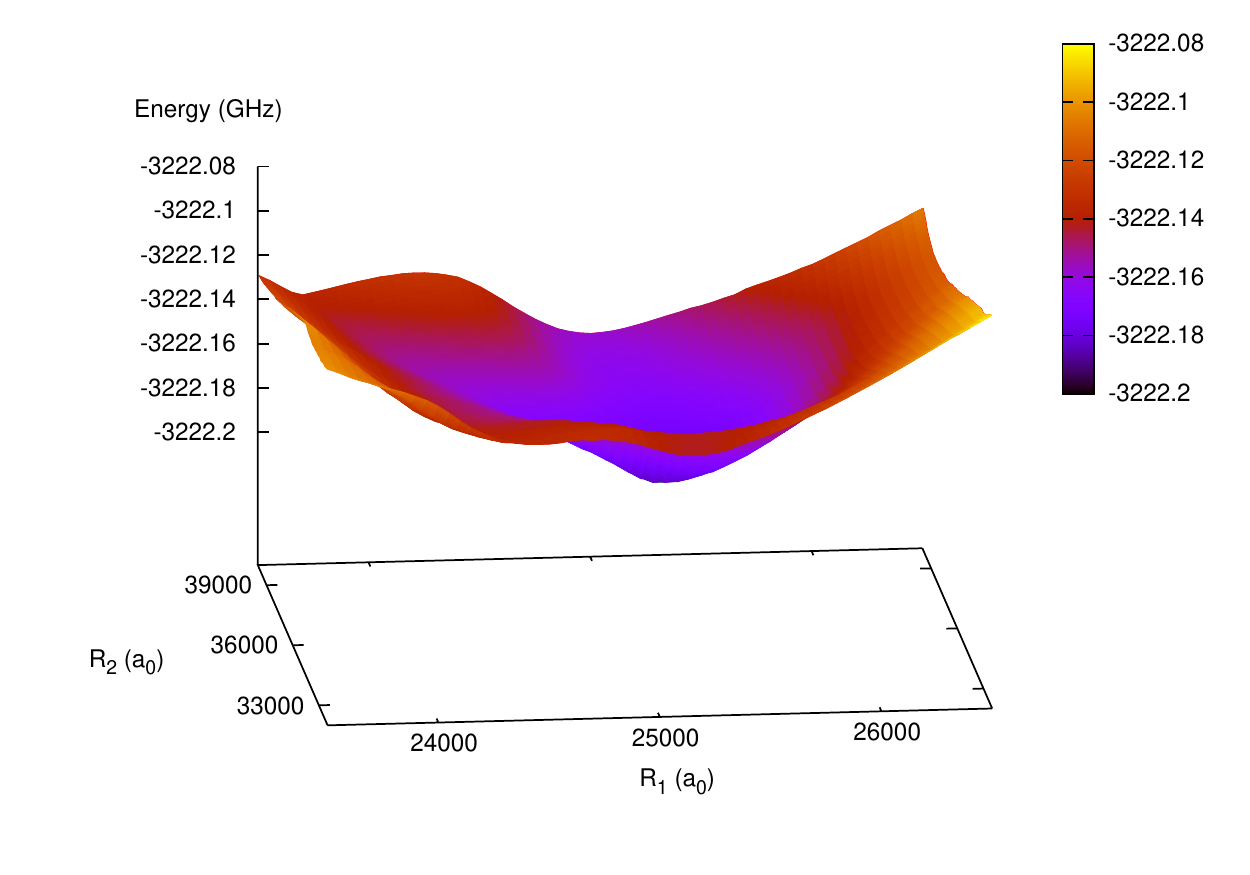}
	\caption{(Color online) Potential surface corresponding to the\\ 
	         $\ket{59s\frac{1}{2}-\frac{1}{2};55d\frac{3}{2}\frac{3}{2};58d\frac{3}{2}-\frac{1}{2}}$ asymptotic state.
	         The color scheme
	         denotes the energy values given in GHz, with the scale presented to the right of
	         the plot.
	         The well exhibited in this surface is 60-100 MHz deep.
	         }
	\label{fig:399}
\end{figure}
%%%%%%%%%%%%%%%%%%%%%%%%%%%%%%%%%%%%%%%%%%%%%%%%%%%%%%%%%%%%%%%
\section{Conclusions}
\label{sec:conc}
The work presented in this letter demonstrates results for the $\Omega=1/2$
symmetry of
$58p+58p+58p$ rubidium Rydberg atoms. During the course of our analysis, we also
examined the surfaces corresponding to asymptotic states near the Rb $58s+58s+58s$ 
asymptote, but no potential wells were found for this case. The formalism could 
also be applied to the $58d+58d+58d$ case, but this would correspond to a large increase
in computation time and thus, is outside the scope of the work shown in this
paper. Due to an increased number of basis states for the triple $58d$ case, we would expect 
to find additional wells, however.

In addition, the methodology presented here can be applied to Rydberg states of other
$\Omega$-symmetry values as well as to other alkali elements. 
The current literature regarding ultracold multi-body Rydberg physics involves 
one Rydberg atom interacting with multiple ground state atoms or a ground state molecule. 

We seek to continue to analyze cases of Rydberg trimers for various alkali elements,
asymptotes, and linear $\Omega$ symmetries, as well as exploring the energy levels corresponding to 
transverse (bending) modes of oscillation. Similarly, we seek to extend the theory to different molecular
configurations, such as triangular systems. 
The detection of such trimer states could have applications
in a variety of research areas, including quantum information processing and exotic, ultracold 
chemistry.

Furthermore, it might be possible to extend the linear
chain to include $N$-Rydberg atoms, although this is purely speculative at this stage.
Such investigations could prove fruitful
in the advancement of ultracold multibody physics.

\acknowledgments
This work was partially supported by the CSBG division of the Department of Energy.


\begin{thebibliography}{22}%
\makeatletter
\providecommand \@ifxundefined [1]{%
 \@ifx{#1\undefined}
}%
\providecommand \@ifnum [1]{%
 \ifnum #1\expandafter \@firstoftwo
 \else \expandafter \@secondoftwo
 \fi
}%
\providecommand \@ifx [1]{%
 \ifx #1\expandafter \@firstoftwo
 \else \expandafter \@secondoftwo
 \fi
}%
\providecommand \natexlab [1]{#1}%
\providecommand \enquote  [1]{``#1''}%
\providecommand \bibnamefont  [1]{#1}%
\providecommand \bibfnamefont [1]{#1}%
\providecommand \citenamefont [1]{#1}%
\providecommand \href@noop [0]{\@secondoftwo}%
\providecommand \href [0]{\begingroup \@sanitize@url \@href}%
\providecommand \@href[1]{\@@startlink{#1}\@@href}%
\providecommand \@@href[1]{\endgroup#1\@@endlink}%
\providecommand \@sanitize@url [0]{\catcode `\\12\catcode `\$12\catcode
  `\&12\catcode `\#12\catcode `\^12\catcode `\_12\catcode `\%12\relax}%
\providecommand \@@startlink[1]{}%
\providecommand \@@endlink[0]{}%
\providecommand \url  [0]{\begingroup\@sanitize@url \@url }%
\providecommand \@url [1]{\endgroup\@href {#1}{\urlprefix }}%
\providecommand \urlprefix  [0]{URL }%
\providecommand \Eprint [0]{\href }%
\providecommand \doibase [0]{http://dx.doi.org/}%
\providecommand \selectlanguage [0]{\@gobble}%
\providecommand \bibinfo  [0]{\@secondoftwo}%
\providecommand \bibfield  [0]{\@secondoftwo}%
\providecommand \translation [1]{[#1]}%
\providecommand \BibitemOpen [0]{}%
\providecommand \bibitemStop [0]{}%
\providecommand \bibitemNoStop [0]{.\EOS\space}%
\providecommand \EOS [0]{\spacefactor3000\relax}%
\providecommand \BibitemShut  [1]{\csname bibitem#1\endcsname}%
\let\auto@bib@innerbib\@empty
%</preamble>
\bibitem [{\citenamefont {Gallagher}(1994)}]{Gallagher}%
  \BibitemOpen
  \bibfield  {author} {\bibinfo {author} {\bibfnamefont {T.}~\bibnamefont
  {Gallagher}},\ }\href@noop {} {\emph {\bibinfo {title} {Rydberg Atoms}}}\
  (\bibinfo  {publisher} {Cambridge University Press},\ \bibinfo {address}
  {Cambridge, United Kingdom},\ \bibinfo {year} {1994})\BibitemShut {NoStop}%
\bibitem [{\citenamefont {Anderson}\ \emph {et~al.}(1998)\citenamefont
  {Anderson}, \citenamefont {Veale},\ and\ \citenamefont
  {Gallagher}}]{Anderson}%
  \BibitemOpen
  \bibfield  {author} {\bibinfo {author} {\bibfnamefont {W.~R.}\ \bibnamefont
  {Anderson}}, \bibinfo {author} {\bibfnamefont {J.~R.}\ \bibnamefont {Veale}},
  \ and\ \bibinfo {author} {\bibfnamefont {T.~F.}\ \bibnamefont {Gallagher}},\
  }\href {\doibase 10.1103/PhysRevLett.80.249} {\bibfield  {journal} {\bibinfo
  {journal} {Phys. Rev. Lett.}\ }\textbf {\bibinfo {volume} {80}},\ \bibinfo
  {pages} {249} (\bibinfo {year} {1998})}\BibitemShut {NoStop}%
\bibitem [{\citenamefont {Mourachko}\ \emph {et~al.}(1998)\citenamefont
  {Mourachko}, \citenamefont {Comparat}, \citenamefont {de~Tomasi},
  \citenamefont {Fioretti}, \citenamefont {Nosbaum}, \citenamefont {Akulin},\
  and\ \citenamefont {Pillet}}]{Mourachko}%
  \BibitemOpen
  \bibfield  {author} {\bibinfo {author} {\bibfnamefont {I.}~\bibnamefont
  {Mourachko}}, \bibinfo {author} {\bibfnamefont {D.}~\bibnamefont {Comparat}},
  \bibinfo {author} {\bibfnamefont {F.}~\bibnamefont {de~Tomasi}}, \bibinfo
  {author} {\bibfnamefont {A.}~\bibnamefont {Fioretti}}, \bibinfo {author}
  {\bibfnamefont {P.}~\bibnamefont {Nosbaum}}, \bibinfo {author} {\bibfnamefont
  {V.~M.}\ \bibnamefont {Akulin}}, \ and\ \bibinfo {author} {\bibfnamefont
  {P.}~\bibnamefont {Pillet}},\ }\href {\doibase 10.1103/PhysRevLett.80.253}
  {\bibfield  {journal} {\bibinfo  {journal} {Phys. Rev. Lett.}\ }\textbf
  {\bibinfo {volume} {80}},\ \bibinfo {pages} {253} (\bibinfo {year}
  {1998})}\BibitemShut {NoStop}%
\bibitem [{\citenamefont {Saffman}\ \emph {et~al.}(2010)\citenamefont
  {Saffman}, \citenamefont {Walker},\ and\ \citenamefont
  {M\o{}lmer}}]{Saffman-RMP}%
  \BibitemOpen
  \bibfield  {author} {\bibinfo {author} {\bibfnamefont {M.}~\bibnamefont
  {Saffman}}, \bibinfo {author} {\bibfnamefont {T.~G.}\ \bibnamefont {Walker}},
  \ and\ \bibinfo {author} {\bibfnamefont {K.}~\bibnamefont {M\o{}lmer}},\
  }\href {\doibase 10.1103/RevModPhys.82.2313} {\bibfield  {journal} {\bibinfo
  {journal} {Rev. Mod. Phys.}\ }\textbf {\bibinfo {volume} {82}},\ \bibinfo
  {pages} {2313} (\bibinfo {year} {2010})}\BibitemShut {NoStop}%
\bibitem [{\citenamefont {Greene}\ \emph {et~al.}(2000)\citenamefont {Greene},
  \citenamefont {Dickinson},\ and\ \citenamefont {Sadeghpour}}]{trilobites}%
  \BibitemOpen
  \bibfield  {author} {\bibinfo {author} {\bibfnamefont {C.~H.}\ \bibnamefont
  {Greene}}, \bibinfo {author} {\bibfnamefont {A.~S.}\ \bibnamefont
  {Dickinson}}, \ and\ \bibinfo {author} {\bibfnamefont {H.~R.}\ \bibnamefont
  {Sadeghpour}},\ }\href {\doibase 10.1103/PhysRevLett.85.2458} {\bibfield
  {journal} {\bibinfo  {journal} {Phys. Rev. Lett.}\ }\textbf {\bibinfo
  {volume} {85}},\ \bibinfo {pages} {2458} (\bibinfo {year}
  {2000})}\BibitemShut {NoStop}%
\bibitem [{\citenamefont {Bendowsky}\ \emph {et~al.}(2009)\citenamefont
  {Bendowsky}, \citenamefont {Butscher}, \citenamefont {Nipper}, \citenamefont
  {Shaffer}, \citenamefont {L\"ow},\ and\ \citenamefont {Pfau}}]{pfau}%
  \BibitemOpen
  \bibfield  {author} {\bibinfo {author} {\bibfnamefont {V.}~\bibnamefont
  {Bendowsky}}, \bibinfo {author} {\bibfnamefont {B.}~\bibnamefont {Butscher}},
  \bibinfo {author} {\bibfnamefont {J.}~\bibnamefont {Nipper}}, \bibinfo
  {author} {\bibfnamefont {J.~P.}\ \bibnamefont {Shaffer}}, \bibinfo {author}
  {\bibfnamefont {R.}~\bibnamefont {L\"ow}}, \ and\ \bibinfo {author}
  {\bibfnamefont {T.}~\bibnamefont {Pfau}},\ }\href {\doibase
  10.1038/nature07945} {\bibfield  {journal} {\bibinfo  {journal} {Nature}\
  }\textbf {\bibinfo {volume} {458}},\ \bibinfo {pages} {1005} (\bibinfo {year}
  {2009})}\BibitemShut {NoStop}%
\bibitem [{\citenamefont {Boisseau}\ \emph {et~al.}(2002)\citenamefont
  {Boisseau}, \citenamefont {Simbotin},\ and\ \citenamefont
  {C\^ot\'e}}]{macro-old}%
  \BibitemOpen
  \bibfield  {author} {\bibinfo {author} {\bibfnamefont {C.}~\bibnamefont
  {Boisseau}}, \bibinfo {author} {\bibfnamefont {I.}~\bibnamefont {Simbotin}},
  \ and\ \bibinfo {author} {\bibfnamefont {R.}~\bibnamefont {C\^ot\'e}},\
  }\href {\doibase 10.1103/PhysRevLett.88.133004} {\bibfield  {journal}
  {\bibinfo  {journal} {Phys. Rev. Lett.}\ }\textbf {\bibinfo {volume} {88}},\
  \bibinfo {pages} {133004} (\bibinfo {year} {2002})}\BibitemShut {NoStop}%
\bibitem [{\citenamefont {Samboy}\ \emph {et~al.}(2011)\citenamefont {Samboy},
  \citenamefont {Stanojevic},\ and\ \citenamefont {C\^ot\'e}}]{Samboy}%
  \BibitemOpen
  \bibfield  {author} {\bibinfo {author} {\bibfnamefont {N.}~\bibnamefont
  {Samboy}}, \bibinfo {author} {\bibfnamefont {J.}~\bibnamefont {Stanojevic}},
  \ and\ \bibinfo {author} {\bibfnamefont {R.}~\bibnamefont {C\^ot\'e}},\
  }\href {\doibase 10.1103/PhysRevA.83.050501} {\bibfield  {journal} {\bibinfo
  {journal} {Phys. Rev. A}\ }\textbf {\bibinfo {volume} {83}},\ \bibinfo
  {pages} {050501} (\bibinfo {year} {2011})}\BibitemShut {NoStop}%
\bibitem [{\citenamefont {Samboy}\ and\ \citenamefont
  {C\^ot\'e}(2011)}]{Samboy-JpB}%
  \BibitemOpen
  \bibfield  {author} {\bibinfo {author} {\bibfnamefont {N.}~\bibnamefont
  {Samboy}}\ and\ \bibinfo {author} {\bibfnamefont {R.}~\bibnamefont
  {C\^ot\'e}},\ }\href {http://stacks.iop.org/0953-4075/44/i=18/a=184006}
  {\bibfield  {journal} {\bibinfo  {journal} {Journal of Physics B: Atomic,
  Molecular and Optical Physics}\ }\textbf {\bibinfo {volume} {44}},\ \bibinfo
  {pages} {184006} (\bibinfo {year} {2011})}\BibitemShut {NoStop}%
\bibitem [{\citenamefont {Overstreet}\ \emph {et~al.}(2009)\citenamefont
  {Overstreet}, \citenamefont {Schwettmann}, \citenamefont {Tallant},
  \citenamefont {Booth},\ and\ \citenamefont {Shaffer}}]{shaffer-NPHYS}%
  \BibitemOpen
  \bibfield  {author} {\bibinfo {author} {\bibfnamefont {K.~R.}\ \bibnamefont
  {Overstreet}}, \bibinfo {author} {\bibfnamefont {A.}~\bibnamefont
  {Schwettmann}}, \bibinfo {author} {\bibfnamefont {J.}~\bibnamefont
  {Tallant}}, \bibinfo {author} {\bibfnamefont {D.}~\bibnamefont {Booth}}, \
  and\ \bibinfo {author} {\bibfnamefont {J.~P.}\ \bibnamefont {Shaffer}},\
  }\href {\doibase 10.1038/nphys1307} {\bibfield  {journal} {\bibinfo
  {journal} {Nature Physics}\ }\textbf {\bibinfo {volume} {5}},\ \bibinfo
  {pages} {581} (\bibinfo {year} {2009})}\BibitemShut {NoStop}%
\bibitem [{\citenamefont {Byrd}\ \emph {et~al.}(2009)\citenamefont {Byrd},
  \citenamefont {Montgomery}, \citenamefont {Michels},\ and\ \citenamefont
  {C\^ot\'e}}]{Byrd-Li3}%
  \BibitemOpen
  \bibfield  {author} {\bibinfo {author} {\bibfnamefont {J.~N.}\ \bibnamefont
  {Byrd}}, \bibinfo {author} {\bibfnamefont {J.~A.}\ \bibnamefont
  {Montgomery}}, \bibinfo {author} {\bibfnamefont {H.~H.}\ \bibnamefont
  {Michels}}, \ and\ \bibinfo {author} {\bibfnamefont {R.}~\bibnamefont
  {C\^ot\'e}},\ }\href {\doibase 10.1002/qua.22063} {\bibfield  {journal}
  {\bibinfo  {journal} {International Journal of Quantum Chemistry}\ }\textbf
  {\bibinfo {volume} {109}},\ \bibinfo {pages} {3112} (\bibinfo {year}
  {2009})}\BibitemShut {NoStop}%
\bibitem [{\citenamefont {Sold\'an}\ \emph {et~al.}(2002)\citenamefont
  {Sold\'an}, \citenamefont {Cvita\ifmmode~\check{s}\else \v{s}\fi{}},
  \citenamefont {Hutson}, \citenamefont {Honvault},\ and\ \citenamefont
  {Launay}}]{Atom-diatom}%
  \BibitemOpen
  \bibfield  {author} {\bibinfo {author} {\bibfnamefont {P.}~\bibnamefont
  {Sold\'an}}, \bibinfo {author} {\bibfnamefont {M.~T.}\ \bibnamefont
  {Cvita\ifmmode~\check{s}\else \v{s}\fi{}}}, \bibinfo {author} {\bibfnamefont
  {J.~M.}\ \bibnamefont {Hutson}}, \bibinfo {author} {\bibfnamefont
  {P.}~\bibnamefont {Honvault}}, \ and\ \bibinfo {author} {\bibfnamefont
  {J. M.}\ \bibnamefont {Launay}},\ }\href {\doibase
  10.1103/PhysRevLett.89.153201} {\bibfield  {journal} {\bibinfo  {journal}
  {Phys. Rev. Lett.}\ }\textbf {\bibinfo {volume} {89}},\ \bibinfo {pages}
  {153201} (\bibinfo {year} {2002})}\BibitemShut {NoStop}%
\bibitem [{\citenamefont {Parazzoli}\ \emph {et~al.}(2011)\citenamefont
  {Parazzoli}, \citenamefont {Fitch}, \citenamefont {\ifmmode~\dot{Z}\else
  \.{Z}\fi{}uchowski}, \citenamefont {Hutson},\ and\ \citenamefont
  {Lewandowski}}]{Lewandowski11}%
  \BibitemOpen
  \bibfield  {author} {\bibinfo {author} {\bibfnamefont {L.~P.}\ \bibnamefont
  {Parazzoli}}, \bibinfo {author} {\bibfnamefont {N.~J.}\ \bibnamefont
  {Fitch}}, \bibinfo {author} {\bibfnamefont {P.~S.}\ \bibnamefont
  {\ifmmode~\dot{Z}\else \.{Z}\fi{}uchowski}}, \bibinfo {author} {\bibfnamefont
  {J.~M.}\ \bibnamefont {Hutson}}, \ and\ \bibinfo {author} {\bibfnamefont
  {H.~J.}\ \bibnamefont {Lewandowski}},\ }\href {\doibase
  10.1103/PhysRevLett.106.193201} {\bibfield  {journal} {\bibinfo  {journal}
  {Phys. Rev. Lett.}\ }\textbf {\bibinfo {volume} {106}},\ \bibinfo {pages}
  {193201} (\bibinfo {year} {2011})}\BibitemShut {NoStop}%
\bibitem [{\citenamefont {Zemke}\ \emph {et~al.}(2010)\citenamefont {Zemke},
  \citenamefont {Byrd}, \citenamefont {Michels}, \citenamefont {John
  A.~Montgomery},\ and\ \citenamefont {Stwalley}}]{Byrd-NaK}%
  \BibitemOpen
  \bibfield  {author} {\bibinfo {author} {\bibfnamefont {W.~T.}\ \bibnamefont
  {Zemke}}, \bibinfo {author} {\bibfnamefont {J.~N.}\ \bibnamefont {Byrd}},
  \bibinfo {author} {\bibfnamefont {H.~H.}\ \bibnamefont {Michels}}, \bibinfo
  {author} {\bibfnamefont {J.}~\bibnamefont {John A.~Montgomery}}, \ and\
  \bibinfo {author} {\bibfnamefont {W.~C.}\ \bibnamefont {Stwalley}},\ }\href
  {\doibase 10.1063/1.3454656} {\bibfield  {journal} {\bibinfo  {journal} {J.
  Chem. Phys.}\ }\textbf {\bibinfo {volume} {132}} (\bibinfo {year} {2010}),\
  10.1063/1.3454656}\BibitemShut {NoStop}%
\bibitem [{\citenamefont {Byrd}\ \emph {et~al.}(2010)\citenamefont {Byrd},
  \citenamefont {Montgomery},\ and\ \citenamefont {C\^ot\'e}}]{Byrd-KRb}%
  \BibitemOpen
  \bibfield  {author} {\bibinfo {author} {\bibfnamefont {J.~N.}\ \bibnamefont
  {Byrd}}, \bibinfo {author} {\bibfnamefont {J.~A.}\ \bibnamefont
  {Montgomery}}, \ and\ \bibinfo {author} {\bibfnamefont {R.}~\bibnamefont
  {C\^ot\'e}},\ }\href {\doibase 10.1103/PhysRevA.82.010502} {\bibfield
  {journal} {\bibinfo  {journal} {Phys. Rev. A}\ }\textbf {\bibinfo {volume}
  {82}},\ \bibinfo {pages} {010502} (\bibinfo {year} {2010})}\BibitemShut
  {NoStop}%
\bibitem [{\citenamefont {Rittenhouse}\ \emph {et~al.}(2011)\citenamefont
  {Rittenhouse}, \citenamefont {Mayle}, \citenamefont {Schmelcher},\ and\
  \citenamefont {Sadeghpour}}]{Sadeghpour11}%
  \BibitemOpen
  \bibfield  {author} {\bibinfo {author} {\bibfnamefont {S.~T.}\ \bibnamefont
  {Rittenhouse}}, \bibinfo {author} {\bibfnamefont {M.}~\bibnamefont {Mayle}},
  \bibinfo {author} {\bibfnamefont {P.}~\bibnamefont {Schmelcher}}, \ and\
  \bibinfo {author} {\bibfnamefont {H.~R.}\ \bibnamefont {Sadeghpour}},\ }\href
  {http://stacks.iop.org/0953-4075/44/i=18/a=184005} {\bibfield  {journal}
  {\bibinfo  {journal} {Journal of Physics B: Atomic, Molecular and Optical
  Physics}\ }\textbf {\bibinfo {volume} {44}},\ \bibinfo {pages} {184005}
  (\bibinfo {year} {2011})}\BibitemShut {NoStop}%
\bibitem [{\citenamefont {Rittenhouse}\ and\ \citenamefont
  {Sadeghpour}(2010)}]{Sadeghpour10}%
  \BibitemOpen
  \bibfield  {author} {\bibinfo {author} {\bibfnamefont {S.~T.}\ \bibnamefont
  {Rittenhouse}}\ and\ \bibinfo {author} {\bibfnamefont {H.~R.}\ \bibnamefont
  {Sadeghpour}},\ }\href {\doibase 10.1103/PhysRevLett.104.243002} {\bibfield
  {journal} {\bibinfo  {journal} {Phys. Rev. Lett.}\ }\textbf {\bibinfo
  {volume} {104}},\ \bibinfo {pages} {243002} (\bibinfo {year}
  {2010})}\BibitemShut {NoStop}%
\bibitem [{\citenamefont {Liu}\ and\ \citenamefont {Rost}(2006)}]{Rost-poly}%
  \BibitemOpen
  \bibfield  {author} {\bibinfo {author} {\bibfnamefont {I.~C.~H.}\
  \bibnamefont {Liu}}\ and\ \bibinfo {author} {\bibfnamefont {J.~M.}\
  \bibnamefont {Rost}},\ }\href {\doibase 10.1140/epjd/e2006-00098-x}
  {\bibfield  {journal} {\bibinfo  {journal} {Eur. Phys. J. D}\ }\textbf
  {\bibinfo {volume} {40}},\ \bibinfo {pages} {65} (\bibinfo {year}
  {2006})}\BibitemShut {NoStop}%
\bibitem [{\citenamefont {Liu}\ \emph {et~al.}(2009)\citenamefont {Liu},
  \citenamefont {Stanojevic},\ and\ \citenamefont {Rost}}]{Jovica-trimers}%
  \BibitemOpen
  \bibfield  {author} {\bibinfo {author} {\bibfnamefont {I.~C.~H.}\
  \bibnamefont {Liu}}, \bibinfo {author} {\bibfnamefont {J.}~\bibnamefont
  {Stanojevic}}, \ and\ \bibinfo {author} {\bibfnamefont {J.~M.}\ \bibnamefont
  {Rost}},\ }\href {\doibase 10.1103/PhysRevLett.102.173001} {\bibfield
  {journal} {\bibinfo  {journal} {Phys. Rev. Lett.}\ }\textbf {\bibinfo
  {volume} {102}},\ \bibinfo {pages} {173001} (\bibinfo {year}
  {2009})}\BibitemShut {NoStop}%
\bibitem [{\citenamefont {Stanojevic}\ \emph {et~al.}(2006)\citenamefont
  {Stanojevic}, \citenamefont {C\^ot\'e}, \citenamefont {Tong}, \citenamefont
  {Farooqi}, \citenamefont {Eyler},\ and\ \citenamefont {Gould}}]{Jovica}%
  \BibitemOpen
  \bibfield  {author} {\bibinfo {author} {\bibfnamefont {J.}~\bibnamefont
  {Stanojevic}}, \bibinfo {author} {\bibfnamefont {R.}~\bibnamefont
  {C\^ot\'e}}, \bibinfo {author} {\bibfnamefont {D.}~\bibnamefont {Tong}},
  \bibinfo {author} {\bibfnamefont {S.}~\bibnamefont {Farooqi}}, \bibinfo
  {author} {\bibfnamefont {E.}~\bibnamefont {Eyler}}, \ and\ \bibinfo {author}
  {\bibfnamefont {P.}~\bibnamefont {Gould}},\ }\href {\doibase
  10.1140/epjd/e2006-00143-x} {\bibfield  {journal} {\bibinfo  {journal} {Eur.
  Phys. J. D}\ }\textbf {\bibinfo {volume} {40}},\ \bibinfo {pages} {3}
  (\bibinfo {year} {2006})}\BibitemShut {NoStop}%
\bibitem [{\citenamefont {LeRoy}(1974)}]{LeRoy}%
  \BibitemOpen
  \bibfield  {author} {\bibinfo {author} {\bibfnamefont {R.~J.}\ \bibnamefont
  {LeRoy}},\ }\href@noop {} {\bibfield  {journal} {\bibinfo  {journal} {Can. J.
  Phys.}\ }\textbf {\bibinfo {volume} {52}},\ \bibinfo {pages} {246} (\bibinfo
  {year} {1974})}\BibitemShut {NoStop}%
\bibitem [{\citenamefont {Beterov}\ \emph {et~al.}(2009)\citenamefont
  {Beterov}, \citenamefont {Ryabtsev}, \citenamefont {Tretyakov},\ and\
  \citenamefont {Entin}}]{RydLife}%
  \BibitemOpen
  \bibfield  {author} {\bibinfo {author} {\bibfnamefont {I.~I.}\ \bibnamefont
  {Beterov}}, \bibinfo {author} {\bibfnamefont {I.~I.}\ \bibnamefont
  {Ryabtsev}}, \bibinfo {author} {\bibfnamefont {D.~B.}\ \bibnamefont
  {Tretyakov}}, \ and\ \bibinfo {author} {\bibfnamefont {V.~M.}\ \bibnamefont
  {Entin}},\ }\href {\doibase 10.1103/PhysRevA.79.052504} {\bibfield  {journal}
  {\bibinfo  {journal} {Phys. Rev. A}\ }\textbf {\bibinfo {volume} {79}},\
  \bibinfo {pages} {052504} (\bibinfo {year} {2009})}\BibitemShut {NoStop}%
\end{thebibliography}
\end{document}